\RequirePackage{ifpdf}
\ifpdf 
\documentclass[pdftex]{sigma}
\else
\documentclass{sigma}
\fi

\numberwithin{equation}{section}

\begin{document}

\allowdisplaybreaks

\renewcommand{\thefootnote}{$\star$}

\renewcommand{\PaperNumber}{063}

\FirstPageHeading

\ShortArticleName{Balance Systems and the Variational Bicomplex}

\ArticleName{Balance Systems and the Variational Bicomplex\footnote{This paper is a
contribution to the Special Issue ``Symmetry, Separation, Super-integrability and Special Functions~(S$^4$)''. The
full collection is available at
\href{http://www.emis.de/journals/SIGMA/S4.html}{http://www.emis.de/journals/SIGMA/S4.html}}}

\Author{Serge PRESTON}

\AuthorNameForHeading{S.~Preston}

\Address{Department of Mathematics and Statistics, Portland State University,\\
Portland, OR, 97207-0751, USA}
\Email{\href{mailto:serge@pdx.edu}{serge@pdx.edu}}
\URLaddress{\url{http://www.mth.pdx.edu/~serge/}}

\ArticleDates{Received January 27, 2011, in f\/inal form June 30, 2011;  Published online July 09, 2011}

\Abstract{In this work we show that the systems of balance equations (balance systems) of continuum thermodynamics occupy a natural place in the variational bicomplex for\-ma\-lism.  We apply the vertical homotopy decomposition to get a local splitting (in a convenient domain) of a general balance system as the sum of a Lagrangian part and a complemental ``pure non-Lagrangian'' balance system.  In the case when derivatives of the dynamical f\/ields do not enter the constitutive relations of the balance system, the ``pure non-Lagrangian'' systems coincide with the systems introduced by S.~Godunov [{\it Soviet Math. Dokl.} {\bf 2} (1961), 947--948] and, later, asserted as the canonical hyperbolic form of balance systems in [M\"uller~I., Ruggeri~T.,
Rational extended thermodynamics, 2nd ed., {\it Springer Tracts in Natural Philosophy}, Vol.~37, Springer-Verlag, New York, 1998].}

\Keywords{variational bicomplex; balance equations}

\Classification{49Q99; 35Q80}

 \section{Introduction}

Systems of balance equations (balance systems) are a cornerstone of the mathematical models of continuum
 thermodynamics. That is why the usage of variational methods for the study of such systems was attempted in the corresponding domains of classical physics for a long time. For a review of the present state of this research f\/ield, we refer to the monographs \cite{Sieniutycz,SieniutyczFarkas} and the references therein.  These works use the conventional setting of variational calculus based on the action principle.

   Inf\/initesimal alternatives to the action principle were suggested by A.~Poincar\'e and \'E.~Cartan in mechanics, by C.~Caratheodory, H.~Weyl and Th.~Lepage in f\/ield theory and gradually evolved into a mathematical theory of a rare beauty, see, for instance, \cite{Anderson,FatibeneFrancaviglia,Krupka,Olver}. Thus, it seems natural to investigate and to develop methods of inf\/initesimal variational calculus that can be used in the study of general balance systems.  In the papers \cite{Preston2007,Preston2010} it was proved that any balance system allows a natural realization by an appropriately modif\/ied Cartan form (see Section~\ref{section3}). Recently, I've observed that the balance systems present the largest class of systems of partial dif\/ferential equations that can be realized using the forms of Lepage type on the inf\/inite jet bundle $J^{\infty}(\pi)$ of the conf\/igurational bundle $\pi$ \cite{Preston2011}.

     One of the most successful approaches in variational calculus is the method of variational bicomplex, see \cite{Anderson,Olver}. In this paper we show that general systems of balance equations occupy a natural place in the framework of variational bicomplex.

      In Section~\ref{section2} we introduce the necessary notions and notation. In Section~\ref{section3} the balance systems of order $k$ are def\/ined.  We recall the construction of the modif\/ied Poincar\'e--Cartan forms for these systems, leading to the intrinsic formulation of the system of balance equations. An $(n,1)$-form $K_{C}$ carrying all the constituents of a balance system is introduced.

       In Section~\ref{section4} we show that any balance system is realized by the functional form corresponding to the $(n,1)$-form $K_{C}$.  Higher versions of balance systems are introduced here as well.

         In Section~\ref{section5} we apply the vertical homotopy decomposition to get a local splitting of a~ge\-ne\-ral $(n,1)$-form $K_{C}$ into the Lagrangian part (corresponding to a function we call ``quasi-Lagrangian'') and a complemental ``pure non-Lagrangian'' balance system ($K$-splitting). Some properties of this decomposition are presented, including induced splitting of corresponding functional forms ($F$-splitting). Finally, we obtain the condition on the  form $K_{C}$ to have zero Lagrangian part.

        In Section~\ref{section6} we determine the conditions under which the Lagrangian part of the functional form for a balance systems of order zero (when the derivatives of dynamical f\/ields do not enter the constitutive relation) vanishes.  It is proved that in such a case, a ``pure non-Lagrangian'' system is of the type introduced by S.~Godunov in 1961~\cite{Godunov1961} and later asserted by T.~Ruggeri and his coauthors, as the canonical (symmetrical hyperbolic) form of balance systems in rational extended thermodynamics~\cite{MullerRuggeri}. We calculate the ``quasi-Lagrangian'' for the balance systems of hyperelasticity and the 2-dimensional ideal plasticity and show that its input into the corresponding balance system reduces to the source terms in the balance laws.

\section{Settings and notations}\label{section2}

Throughout this paper $\pi:Y\rightarrow X$ is a (\emph{configurational}) f\/ibre bundle with a $n$-dimensional connected paracompact smooth $C^\infty $-manifold $X$ as the base and the total space $Y$, $\dim(Y)=n+m$. The f\/iber of the bundle $\pi$ is a $m$-dimensional connected smooth manifold~$U$.

The base manifold $X$ is endowed with a pseudo-Riemannian metric $G$.  The volume form of metric $G$ will be denoted by $\eta$. Our considerations are mostly local and, as a result, we shall not deal with the boundary of $X$.

We will be using f\/ibred charts $(W, x^\mu, y^i)$ in a domain $W\subset  Y$. Here $(\pi(W),x^\mu )$ is the chart in~$X$ and $y^i$ are coordinates along the f\/ibers.  A local frame corresponding to the chart $(W, x^\mu, y^i)$ will be denoted by $(\partial_{\mu}=\partial_{x^\mu}$, $\partial_{i}=\partial_{y^i})$ (the shorter notation will be used in more cumbersome calculations) and the corresponding coframe   $(dx^\mu ,dy^i)$.

  We introduce the contracted forms $\eta_{\mu}=i_{\partial_{x^\mu}}\eta$. Below we will be using following relations for the forms  $\eta_{\nu}$ without references (here and below $\lambda_{G}=\ln(\sqrt{\vert G\vert})$)
\begin{gather*}dx^\nu \wedge \eta_{\mu}=\delta^{\nu}_{\mu}\eta,\qquad
d\eta_{\mu}=\lambda_{G,\mu}\eta.
\end{gather*}

Section  $s:V \rightarrow Y$ of the bundle  $\pi$, def\/ined in a domain $V \subset X $, corresponds to the collection $\{ y^{i}(x)\}$ of classical f\/ields. Usually these f\/ields are components of tensor f\/ields or tensor densities f\/ields (sections of ``natural bundles''~\cite{FatibeneFrancaviglia}).

For a manifold $M$ we will denote by $\tau_{M}:T(M)\rightarrow M$ the tangent bundle of a manifold $M$ and by $\tau^{*}_{M}:T^{*}(M)\rightarrow M$   its cotangent bundle. For a bundle $Y\rightarrow X$, denote by $\pi_{V}:V(\pi)\rightarrow Y$ the subbundle of tangent bundle $T(Y)$ consisting of the vertical vectors i.e.\ vectors $\xi\in T_{y}(Y)$ such that $\pi_{*y}\xi =0$.

Denote by $\Omega^{r}(M)\rightarrow M$ the bundle of exterior $r$-forms on a manifold $M$ and by  $\Omega^{*}(M)= \oplus_{r=0}^{\infty}  \Omega^{r}(M)$   the dif\/ferential algebra of exterior forms on the manifold~$M$.

Given a f\/iber bundle $\pi:Y\rightarrow X$ denote by $J^{k}(\pi)$ the $k$-jet bundle of sections of the bundle~$\pi$~\cite{KolarMichorSlovak,Saunders}. For a section $s:U\rightarrow Y$, denote by $j^{k}s:U\rightarrow J^{k}(\pi)$ the $k$-jet of section~$s$.

  Denote by $\pi_{kr}:J^{k}(\pi)\rightarrow J^{r}(\pi)$, $k\geq r\geq 0$ the natural projections between the jet bundles of dif\/ferent order and by $\pi_{k}:J^{k}(\pi)\rightarrow X$ the projection to the base manifold $X$.

 Projection mappings $\pi_{k(k-1)}: J^{k}(\pi)\rightarrow J^{k-1}(\pi)$ form the tower of $k$-jet bundles
\begin{gather}  \cdots \rightarrow J^{k}(\pi)\rightarrow J^{k-1}(\pi)\rightarrow \cdots \rightarrow  Y \rightarrow X.  \label{tower}
\end{gather}

A bundle $\pi_{k(k-1)}: J^{k}(\pi) \rightarrow
J^{k-1}(\pi)$
 has the structure of an \emph{affine bundles} modeled on the vector bundle $S^{k} T^{*}(X)\otimes_{\rightarrow J^{k-1}(\pi)} V(\pi)\rightarrow J^{k-1}(\pi)$.

Denote by $J^{\infty}(\pi)$ the inf\/inite jet bundle of bundle $\pi$ -- the inverse limit of the projective sequence \eqref{tower}~\cite{Krupka,Olver,Saunders}. Space $J^{\infty}(\pi)$ is endowed with the structure of inverse limit of dif\/ferentiable manifolds with the natural sheaves of vector f\/ields, dif\/ferential forms etc., making the projections $\pi_{\infty k}:J^{\infty}(\pi)\rightarrow J^{k}(\pi)$ the smooth surjections.  See \cite{KolarMichorSlovak,Krasil'shchikVinogradov,Saunders} for more about structure and properties of $k$-jet bundles.

For a mutliindex $I =\{i_{1},\ldots ,i_{n}\}$, $i_{k}\in \mathbb{N}$ denote by $\partial^{I}$ the dif\/ferential operator $\partial^{I}f=\partial^{i_{1}}_{x^1}  \cdots \partial^{i_{n}}_{x^n}$ in $C^{\infty}(X)$.  To every f\/ibred chart $(W, x^\mu ,y^i )$ in $Y$
there corresponds the f\/ibred chart $(x^\mu$, $y^i$, $z^{i}_{\mu}$, $\vert I \vert =\sum_{s}i_{s}\leq k)$ in the domain $W^k=\pi_{k0}^{-1}(W)\subset J^{k}(\pi)$.  Coordinates $z^{i}_{\Lambda}$ in the f\/ibers of jet bundles are def\/ined by the condition $ z^{\mu}_{I}(j^{k}_{x}s)=\partial^I s^{\mu}(x).$

For $k=1,\ldots,\infty$ the space $J^{k}(\pi)\rightarrow X$ of k-jet bundle is endowed with the \textit{contact $($Cartan$)$ distribution} $Ca^{k}$ def\/ined by the basic \textit{contact forms}
\begin{gather*}
\omega^{i}=dy^i -z^{i}_{\mu}dx^\mu, \qquad \ldots,\qquad \omega^{i}_{I}=dz^{i}_{I}-z^{i}_{I +1_{\nu}}dx^\nu,\qquad \vert I\vert <k.
\end{gather*}
 These forms generate the contact ideal $C^k  \subset \Omega^{*}(J^{k}(\pi))$ in the algebra of all exterior forms. A $p$-form is called $l$-contact if it belongs to the $l$-th degree of this ideal $(C^k)^l \subset \Omega^{*}(J^{k}(\pi)).$  0-contact exterior forms are also called \textit{horizontal $($or $\pi_{k}$-horizontal$)$ forms} (or, sometimes, \emph{semi-basic forms}).

Any exterior $l$-form $\omega^l$ on the space $\Omega^{\infty}(\pi)$ has the decomposition $\omega^l=\sum_{s+r=l}\omega^{(s,r)}$ into its $s$-horizontal, $r$-vertical components $\omega^{(s,r)}$  ($(s,r)$-forms).  Denote by $\Omega^{(s,r)}(J^{\infty}(\pi))$ the space of all $(s,r)$-forms on $J^{\infty}(\pi)$.

Bellow we will use the presentation of the exterior dif\/ferential $d$ as the sum of \textit{horizontal and vertical differentials} $d_H$, $d_V$: for a $q$-form $\nu \in \Omega^{*}(J^{k}(\pi))$ its dif\/ferential $d\nu$ lifted into the $J^{k+1}(\pi)$ is presented as the sum of horizontal and contact (vertical) terms: $d\nu =d_{H}\nu+d_{V}\nu$, see~\cite{Krupka,Olver}. Operators $d_H$, $d_V$ are naturally def\/ined in the space  $\Omega^{*}(J^{\infty})(\pi))$.

We recall that these operators have the following homology properties
\[
d^{2}_{h}=d^{2}_{V}=d_{V}d_{H}+d_{H}d_{V}=0.
\]

In particular, for a function $f\in C^{\infty}(J^{\infty}(\pi))$ (depending on the jet variables $z^{i}_{I}$ up to some degree, say $\vert I\vert \leq k$),
\begin{gather}
df=(d_{\mu}f)dx^\mu +\sum_{\vert I\vert \geq 0}f_{,z^{i}_{I}}\omega^{i}_{I }, \label{2.3}
\end{gather}
where
\begin{gather}
d_{\mu}f =\partial_{x^\mu}f+\sum_{ \vert I\vert \geq 0} z^{i}_{I +1_{\mu}}\partial_{z^{i}_{I}}f \label{2.4}
\end{gather}
is the \textit{total derivative} of the function $f(x,y,z)$ by $x^\mu$.  The series in the formulas \eqref{2.3}, \eqref{2.4} contains f\/inite number of terms: $\vert I\vert \leq k$.

\section[Balance systems, form $K_C$ and the Helmholtz coindition]{Balance systems, form $\boldsymbol{K_C}$ and the Helmholtz coindition}\label{section3}

Let $\pi:Y\rightarrow X^n$ be a conf\/igurational bundle  and $G$ a pseudo-Riemannian metric in $X$.
Balance systems of order $k$ is the system of equations for the f\/ields $y^i (x)$ of the form
\begin{gather}
d_{\mu}F^{\mu}_{i}+F^{\mu}_{i}\lambda_{G,\mu}=\Pi_{i},\qquad i=1,\ldots, m. \label{Bsys}
\end{gather}
Here $x^\mu$, $\mu=0,\ldots,n$ are local coordinates in $X$ -- part of the f\/ibred chart $(W,x^\mu ,y^i)$  in $Y$, $\lambda_{G}=\ln(\sqrt{\vert G\vert})$. Densities $F^{0}_{i}$, f\/luxes $F^{A}_{i}$, $A=1,\ldots,n$ and source/production terms $\Pi_{i}$ are smooth functions on $J^{k}(\pi)$, def\/ined by the constitutive relations $C$, which determines the properties of the physical system (solid, f\/luid, gas, etc.).

In the works \cite{Preston2007,Preston2010} we've associated with a balance systems \eqref{Bsys} the $n+(n+1)$-form (modif\/ied \emph{Cartan form})
\[\Theta_{C}=F^{\mu}_{i}\omega^i \wedge \eta_{\mu}+\Pi_{i}\omega^{i}\wedge \eta+F^{\mu}_{i}\omega^{i}_{\mu}\wedge \eta .\]

The Euler--Lagrange equations, corresponding to this form $\Theta_{C}$
\[
\big(j^{k+1}s\big)^{*}i_{\xi}{\tilde d}\Theta_{C}=0,\qquad \forall \; \xi \in \mathcal{X}\big(J^{k}(\pi)\big),
\]
coincide with the balance system \eqref{Bsys}. Here $\tilde d$ is the ``con''-dif\/ferential acting on a $k+(k+1)$-form $\alpha^k +\beta^{k+1}$ as follows: ${\tilde d}(\alpha^k +\beta^{k+1})=(-d\alpha +\beta)+d\beta$ (see \cite{Preston2007,Preston2010}).

  Now we restrict our consideration to the case of a balance system \eqref{Bsys} of order one (i.e.~$F^{\mu}_{i}$, $\Pi_{i}\in C^{\infty}(J^{1}(\pi))$) and associate with such system the following (1-contact) $(n,1)$-form on $J^{2}(\pi)$
\begin{gather}
K_{C}=\big(F^{\mu}_{i}\omega^{i}_{\mu}+\Pi_{i}\omega^i \big)\wedge \eta. \label{K}
\end{gather}

Next we obtain the following version of (local) Helmholtz condition (comp.~\cite{Anderson}) for a balance system of order one to be Euler--Lagrange system for some Lagrangian $L\in C^{\infty}(J^{1}(\pi))$.

\begin{proposition}\label{proposition1} For a constitutive relation $C$ of order $1$ the following properties are equivalent:
\begin{enumerate}\itemsep=0pt
\item[$1)$] form $K_{C}$ is closed: $dK_{C}=0$,
\item[$2)$] $C$ is locally Lagrangian constitutive relation: $F^{\mu}_{i}=\frac{\partial L}{\partial z^{i}_{\mu}}$, $\Pi_{i}=\frac{\partial L}{\partial y^i}$ for the locally defined function $L\in C^{\infty}(J^{1}(\pi))$.
\end{enumerate}
\end{proposition}

\begin{proof}
Let the form $K_{C}$ be closed.  Calculate the dif\/ferential of this form:
\begin{gather*}
dK_{C}=\big[(\partial_{y^j}\Pi_{i})\omega^{j}\wedge \omega^{i}+(\partial_{y^i}F^{\mu}_{j}-\partial_{z^{j}_{\mu}}\Pi_i )\omega^{i}\wedge \omega^{j}_{\mu} +(\partial_{z^{j}_{\nu}}F^{\mu}_{i})\omega^{j}_{\nu}\wedge \omega^{i}_{\mu} \big]\wedge \eta  \\
\phantom{dK_{C}}{}
=\big[(\partial_{y^j}\Pi_{i})dy^{j}\wedge dy^{i}+(\partial_{y^i}F^{\mu}_{j}-\partial_{z^{j}_{\mu}}\Pi_i )dy^{i}\wedge dz^{j}_{\mu} +(\partial_{z^{j}_{\nu}}F^{\mu}_{i})dz^{j}_{\nu}\wedge dz^{i}_{\mu} \big]\wedge \eta.
\end{gather*}
The second line in last formula shows that if we consider $F^{\mu}_{i}$ and $\Pi_{i}$ as functions of vertical variables $y^j$, $z^{j}_{\nu}$ (i.e.\ if we f\/ix $x$) then the 1-form of \emph{these variables} $F^{\mu}_{i}dz^{i}_{\mu}+\Pi_{i}dy^i $ is closed. Then, locally, by Poincar\'e lemma, $F^{\mu}_{i}dz^{i}_{\mu}+\Pi_{i}dy^i =d_{V}L$ for some function $L(x,y,z)$. $d_V$ here is the dif\/ferential by the variables $y^j$, $z^{j}_{\nu}$.  It is easy to see now that $K_{C}=d(L\eta)$.

Proof that the statement 1) follows from 2) is trivial.
\end{proof}

\begin{remark} Since $dK_{C}=d_{V}K_{C}$, the statement of Proposition~\ref{proposition1}, equivalent to the local equality $K_{C}=d_{V}(L\eta)$, follows from the qualif\/ied exactness of the column of the augmented variational bicomplex (see bellow), starting at $\Omega^{n,0}$, at the term $\Omega^{n,1}$.
\end{remark}

\section{Balance systems and the variational bicomplex}\label{section4}

 Proposition~\ref{proposition1} for a balance system~\eqref{Bsys} to be Lagrangian points at the possibility to interpret and study the balance systems using the variational bicomplex, see~\cite{Anderson,Olver}. In this section we realize this possibility.  In presenting the variational bicomplex we will follow \cite{Anderson}.

The sheaf of $k$-contact $s+k$-forms on the inf\/inite jet bundle $J^{\infty}(\pi)$ is denoted by $\Omega^{s,k}(J^{\infty}(\pi))$ or, for shortness, $\Omega^{s,k}$.

We will be using the \textit{augmented variational bicomplex} -- the diagram where all the columns and rows are complexes of sheaves of dif\/ferential forms (right column is the complex of factor-sheaves $\mathcal{F}^k={\rm Im}(I:\Omega^{n,k}(J^{\infty}(\pi))\rightarrow \Omega^{n,s}(J^{\infty}(\pi))$).
\begin{gather}
\mbox{\footnotesize $\begin{CD}
@. @.  @AA d_{V}A @. @. @AA d_{V}A @AA \delta_{V}A\\
@. 0 @>>>  \Omega^{0,3}  @.  @. \cdot \  \cdot \ \cdot \    @. \Omega^{n,3}  @>I>>  \mathcal{F}^3 @>>> 0\\
@. @. @AA d_{V}A @. @.  @AA d_{V}A @AA \delta_{V}A \\
@. 0 @>>> \Omega^{0,2} @>d_H>> \Omega^{1,2} @>d_H>> \cdot \  \cdot \ \cdot \  \Omega^{n-1,2} @>d_H>> \Omega^{n,2} @>I >>\mathcal{F}^2 @>>> 0\\
@.  @. @Ad_VAA @Ad_VAA @Ad_V AA @Ad_VAA  @A\delta_V AA  \\
@. 0 @>>> \Omega^{0,1} @>d_H>> \Omega^{1,1} @>d_H >> \ \cdot \  \cdot\ \cdot \ \ \Omega^{n-1,1} @>d_H>> \Omega^{n,1} @>I>>\mathcal{F}^1 @>>> 0 \\
@. @.       @Ad_VAA    @Ad_VAA @Ad_VAA  @Ad_VAA  @. \\
0 @>>> \mathbb{R} @>>> \Omega^{0,0} @>d_H>> \Omega^{1,0} @>d_H >> \cdot \  \cdot \ \cdot \  \Omega^{n-1,0} @>d_H>> \Omega^{n,0} @.\\
 \end{CD}$} \hspace*{-10mm}\label{Bik}
\end{gather}

Recall also the Euler--Lagrange mapping $\mathcal{E} $:
\[
\mathcal{E}: \ \Omega^{n,0}\rightarrow \mathcal{F}^1,\qquad \mathcal{E}([L\eta])=I(d_{V}L\eta),
\]
where $I:\Omega^{n,s}(J^{\infty}(\pi))\rightarrow \Omega^{n,s}(J^{\infty}(\pi))$ is the ``interior Euler (homotopy) operator'' (see~\eqref{Eop} below).  Mapping $\mathcal{E}$ is well def\/ined because of the relation $d_{V}\circ d_{H}=-d_{H}\circ d_{V}$ between vertical and horizontal dif\/ferentials. In terms of this mapping the Euler--Lagrange equations for a section $s:X\rightarrow Y$ of the conf\/igurational bundle $\pi$ takes the intrinsic form
\begin{gather*}
\big(j^{\infty}s\big)i_{\xi}\mathcal{E}(L\eta)=0,\qquad \forall\; \xi \in \mathcal{X}\big(J^{\infty}(\pi)\big).
\end{gather*}

 Last row and the last column in the bicomplex \eqref{Bik} form the \textit{Euler--Lagrange sequence} where $\mathcal{E}$ is the Euler--Lagrange mapping
 \[
 0 \rightarrow \mathbb{R} \rightarrow \Omega^{0,0} \xrightarrow{d_H}  \Omega^{1,0}  \xrightarrow{d_H}   \cdots  \
 \Omega^{n-1,0} \xrightarrow{d_H}  \Omega^{n.0} \xrightarrow{\mathcal{E}} \mathcal{F}^1 \xrightarrow{\delta_V} \mathcal{F}^{2} \xrightarrow{\delta_V}  \mathcal{F}^3 \xrightarrow{\delta_V} \cdots.
 \]

We recall (see \cite{Anderson,Olver}) that the interior Euler (\emph{homotopy}) operator
\[
I: \ \Omega^{n,s}(J^{\infty}(\pi)) \rightarrow \mathcal{F}^{s}\subset \Omega^{n,s}(J^{\infty}(\pi))
\]
that has, as its image, the subbundle $\mathcal{F}^s$ of functional forms of vertical degree $s$ is def\/ined by the formula
\begin{gather}
I(\omega) =\frac{1}{s}\omega^i \!\wedge \!\!\left[\!\sum_{(i,\Lambda),\vert \Lambda \vert \geq 0} \!\!\!(-d)_{\Lambda} (i_{\partial_{z^{i}_{\Lambda}}}\omega)\right]\! 
= \frac{1}{s}\omega^i \!\wedge\! \Big[ i_{\partial_{i}}\omega \!-\!d_{\mu}(i_{\partial_{z^{i}_{\mu}}}\omega) \!+\!d_{\mu_1 \mu_2}(i_{\partial_{z^{i}_{\mu_1 \mu_2}}}\omega )\! -\!\cdots \Big].\!\!\!\label{Eop}
\end{gather}

\begin{proposition}
Let $K_\mathcal{C}=\Pi_{i}\omega^{i}\wedge \eta +F^{\mu}_{i}\omega^{i}_{\mu}\wedge \eta$  be the $K$-form corresponding to a constitutive relation $C$ of order $k$, see~\eqref{K}.  Then, the balance system \eqref{Bsys} is equivalent to the requirement that
\begin{gather}
j^{1}s^{*} i_{\xi}I(K_{\mathcal{C}})=0 \qquad \text{for\  all}\  \xi \in \mathcal{X}\big(J^{k}(\pi)\big). \label{Eqinv}
\end{gather}
\end{proposition}
\begin{proof}
\[
I(K_{\mathcal{C}})= \omega^i \wedge \big[\Pi_{i}\eta -d_{\mu} (F^{\mu}_{i}\eta)\big] =
\big[\Pi_{i}  -d_{\mu}F^{\mu}_{i} -F^{\mu}_{i}\lambda_{G,\mu}\big]\omega^{i}\wedge \eta .
\]

Let $\xi \in \mathcal{X}(J^{1}(\pi))$. Apply $i_{\xi}$ to the last equality and take the pullback by the mapping $j^{\infty}s$, $s:U\rightarrow Y$ being a section of the bundle $\pi$ over an open subset $U\subset X$.  Equation \eqref{Eqinv} takes the form
\[
\big(\omega^{i}(\xi)\circ j^{1}s\big) [  d_{\mu}F^{\mu}_{i} +F^{\mu}_{i}\lambda_{G,\mu}-\Pi_{i}]\circ j^{\infty}s   \eta =0.
\]
Fulf\/illment of this equality for all (or for many enough) variational vector f\/ields $\xi$ is equivalent to the statement that $s:U\rightarrow Y$ is the solution of the balance system~\eqref{Bsys}.
\end{proof}

\begin{remark} The result of Proposition~\ref{proposition1} shows that a balance system \eqref{Bsys} is \emph{the most general form of PDEs system} that can be obtained from a form of the type~\eqref{K} using the inf\/initesimal variational equation \eqref{Eqinv} in the formalism of variational bicomplex. Similar result for the systems of PDEs starting from an arbitrary Lepage form is obtained in~\cite{Preston2011}.
\end{remark}

\subsection{Higher order balance systems}\label{section4.1}

Balance systems \eqref{Bsys} have found their natural place in the variational bicomplex \eqref{Bik}.  Applying the mapping $I$ to the form $K_{C}$, contracting it with a variation vector f\/ield and taking the pullback by a 2-jet of a section we restore the balance system~\eqref{Bsys}.

Thus, it seems natural to inquire, what higher order construction corresponds to the form $K$ and which system of PDEs corresponds to this construction. It may lead to the def\/inition of a~canonical form of higher order system of balance equations.

Consider an arbitrary $(n,1)$-form on the $J^{\infty}(\pi)$
\begin{gather}
K=\left(\sum_{\vert \Lambda \vert \geq 0}F^{\Lambda}_{i}\omega^{i}_{\Lambda} \right) \wedge \eta,\label{HOform}
\end{gather}
where $F^{\Lambda}_{i}\in C^{\infty}(J^{\infty}(\pi)).$

Calculate the functional form $I(K)$ assuming that when $\sigma =0$, $d_{\sigma}$ is the identity operator. We get
\begin{gather*}
I(K)=\omega^i \wedge \left[ \sum_{\vert \Sigma \vert \geq 0}(-1)^{\vert \Sigma \vert}d_{\Sigma}\big(i_{\partial_{z^{i}_{\Sigma}}}K \big)\right]
=\omega^i \wedge \left[ \sum_{\vert \Sigma \vert \geq 0}(-1)^{\vert \Sigma \vert}d_{\Sigma}\big(F^{\Sigma}_{i}\eta \big)\right] \\
\phantom{I(K)}{}  =
\omega^i \wedge \left[ \sum_{\vert \Sigma \vert \geq 0}(-1)^{\vert \Sigma \vert}d_{\Sigma}\big(F^{\Sigma}_{i}\vert G\vert\big)\frac{1}{\vert G\vert}\eta  \right].
\end{gather*}
Forming the interior derivative of obtained $(n,1)$-form with a vector f\/ield $\xi \in \mathcal{X}(J^{\infty}(\pi))$ and taking pullback of obtained horizontal $n$-form by the $\infty$-jet $j^{\infty}s$ of a section $s:U\rightarrow Y$ we get the system of equations for section $s$
\[
\big(j^{\infty}s\big)^{ *} \sum_{\vert \Sigma \vert \geq 0}(-1)^{\vert \Sigma \vert}\omega^{i}(\xi)d_{\Sigma}\big(F^{\Sigma}_{i}\vert G\vert\big)\frac{1}{\vert G\vert}\eta=0, \qquad i=1,\ldots,m.
\]
Role of the source term in these $m$ equations is played by the functions $F^{0}_{i}$.

Thus, we have proved the following
\begin{proposition} For a $(n,1)$-form $K$ of the form \eqref{HOform} on the infinite jet space $J^{\infty}(\pi)$  and a~section $s:U\rightarrow Y$ where $U\subset X$ is an open subset of $X$, the following statements are equivalent
\begin{enumerate}\itemsep=0pt
\item[$1)$] for all (or, what is equivalent, for $m$ linearly independent)  vector fields $\xi_j$ on $J^{\infty}(\pi)$ such that $\det(\omega^{i}(\xi_j ))\ne 0$,
\[
j^{1}s^{*} i_{\xi}I(K)=0,
\]
\item[$2)$] section $s$ is the solution of the system of ``higher order balance equations''
\begin{gather}
\sum_{\vert \Sigma \vert > 0}\frac{(-1)^{\vert \Sigma \vert-1}}{\vert G\vert} d_{\Sigma}\big(F^{\Sigma}_{i}\vert G\vert\big)=F_{i},\qquad i=1,\ldots, m.\label{HOeq}
\end{gather}
\end{enumerate}
\end{proposition}

\begin{example} Consider the form \eqref{HOform} with $F^{\Lambda}_{i}=\frac{\partial L}{\partial z^{i}_{\Lambda}}$ for a function $L\in C^{\infty}(J^{\infty}(\pi)).$  Then, the system of PDEs \eqref{HOeq} takes the form
\[
\frac{\partial L}{\partial y^i}+\sum_{\vert \Sigma \vert > 0}\frac{1}{\vert G\vert} (-d)_{\Sigma}\left(\frac{\partial L}{\partial z^{i}_{\Sigma}}\vert G\vert\right)=0, \qquad i=1,\ldots, m.
\]
Here $(-d)_{\Sigma}=(-1)^{\Sigma}d_{\Sigma}.$   For $\vert G\vert=1$ this system coincide with the higher order Euler--Lagrange equations, \cite[equation~(3.27)]{CamposdeLeonDiegoVankerschaver} or \cite[Section~4.1]{Olver}.
\end{example}

\section{Vertical splitting of a balance system}\label{section5}

\subsection{Vertical homotopy operators}\label{section5.1}

Recall the construction of the vertical homotopy operator $h^{r,s}_{V}:\Omega^{r,s}\rightarrow \Omega^{r,s-1}$ def\/ined by I.~Anderson~\cite{Anderson}:
\begin{gather}
h^{r,s}_{V}(\omega) =\int_{0}^{1}\frac{1}{t}\Phi^{*}_{\ln(t)}(i_{{\rm pr}\, R}\omega )dt, \label{Vhom}
\end{gather}
where $R=y^i \partial_{y^i}$ is the Liouville (vertical radial) vector f\/ield and
\[
{\rm pr}\, R=y^i \partial_{y^i}+z^{i}_{\mu}\partial_{z^{i}_{\mu}}+\cdots + z^{i}_{\Lambda}\partial_{z^{i}_{\Lambda}}+\cdots
\]
is its prolongation to $J^{\infty}(\pi)$. The f\/low of this vector f\/ield on $J^{\infty}(\pi)$ is the 1-parameter family of dif\/feomorphisms of $J^{\infty}(\pi)$
\[
\Phi_{\epsilon}(x,y,z)=\big(z;e^{\epsilon}y,e^{\epsilon}z\big).
\]
It is proved in \cite{Anderson} that for any $\omega \in \Omega^{r,s}$,
\begin{gather}
\omega =d_{V}[h^{r,s}_{V}(\omega)]+h^{r,s+1}_{V}(d_{V}\omega ). \label{Vdec}
\end{gather}
In the proof of this formula it was shown  \cite[equation~(4.10)]{Anderson} that the integrand  in the formula~\eqref{Vhom} is equal to
\[
\left( \frac{1}{t}\Phi^{*}_{\ln(t)}(i_{{\rm pr}\, R}\omega )\right)[x,y,z]=t^{s-2}(i_{{\rm pr}\, R}\omega )[x;t y,t z ]=t^{s-1}i_{{\rm pr}\, R}\omega[x;t y,t z ].
\]
Here $\omega[x;t y,t z ]$ is the form obtained by evaluating the coef\/f\/icients of the form $\omega$ at the point $[x;t y,t z ]$. Using this in \eqref{Vhom} we get
\begin{gather}
h^{r,s}_{V}(\omega) =\int_{0}^{1}t^{s-1}i_{{\rm pr} R}\omega[x;t y,t z ]dt. \label{Vhom2}
\end{gather}

\subsection[Decomposition of the form $K_{C}$, quasi-Lagrangian]{Decomposition of the form $\boldsymbol{K_{C}}$, quasi-Lagrangian}\label{section5.2}

We will use the formula \eqref{Vhom2} in the cases $r=n$, $s=2,1$.

Let $\omega =K_{C}=\Pi_{i}\omega^{i}\wedge \eta +F^{\mu}_{i}\omega^{i}_{\mu}\wedge \eta \in \Omega^{n,1}$.  Then,
\begin{gather*}
i_{{\rm pr}\, R}\omega[x;t y,t z ]= \Pi_{i}(x,ty,tz)i_{{\rm pr}\, R}\big(\omega^{i}\wedge \eta\big)+F^{\mu}_{i}(x,ty,tz)i_{{\rm pr}\, R}\big(\omega^{i}_{\mu}\wedge \eta\big)\\
\phantom{i_{{\rm pr}\, R}\omega[x;t y,t z ]}{} =\big[\Pi_{i}(x,ty,tz)y^i \eta +F^{\mu}_{i}(x,ty,tz)z^{i}_{\mu}\big]\eta,
\end{gather*}
and $h^{n,1}_{V}K_{C}[x; y, z ]=\tilde L \eta $, where the coef\/f\/icient
\begin{gather} \label{qL}
 {\tilde L}(x,y,z) =\int^{1}_{0}\big[y^i \Pi_{i}(x,ty,tz)dt   +z^{i}_{\mu}F^{\mu}_{i}(x,ty,tz)\big]dt
\end{gather}
of the volume form $\eta$ will be called the ``quasi-Lagrangian'' of the form $K_C$ in the sense of decomposition~\eqref{Vdec}.

Applying the vertical dif\/ferential to $h^{n,1}_{V}K_{C}[x; y, z ]$ we f\/ind the ``Lagrangian part'' of the form~$K_C$
\begin{gather}
d_V h^{n,1}_{V}K_{C}[x; y, z ]= \int_{0}^{1}\left( \Pi_{k}[]+ty^i\Pi_{i,y^k}[]+tz^{i}_{\mu}F^{\mu}_{i,y^k}[]\right) dt  \omega^{k}\wedge \eta\nonumber\\
\phantom{d_V h^{n,1}_{V}K_{C}[x; y, z ]=}{}
+ \int_{0}^{1}\left( ty^i\Pi_{i,z^{j}_{\nu}}[] +F^{\nu}_{j}[] +tz^{i}_{\mu}F^{\mu}_{i,z^{j}_{\nu}}[]\right) dt  \omega^{j}_{\nu}\wedge \eta\nonumber\\
\phantom{d_V h^{n,1}_{V}K_{C}[x; y, z ]=}{}
+\sum_{\Lambda\vert \vert \Lambda\vert >1}\int_{0}^{1}\left( ty^i\Pi_{i,z^{j}_{\Lambda}}[]+tz^{i}_{\mu}F^{\mu}_{i,z^{j}_{\Lambda}}[]\right) dt \omega^{j}_{\Lambda}\wedge \eta. \label{Lag}
\end{gather}

To shorten the formulas we will use the sign $[]$ after function saying that the arguments of this function are $(x,ty^i,tz^{i}_{\mu},\ldots, tz^{i}_{\Sigma})$.

Let $\omega =A^{\Lambda \Sigma}_{ij}\omega^{i}_{\Lambda}\wedge \omega^{j}_{\Sigma}\wedge \eta \in \Omega^{n,2}$ be any $(n,2)$-form.  Then
the integrand in \eqref{Vhom} is equal to
\begin{gather*}
t i_{{\rm pr}\, R}\omega[x,ty,tz]=ti_{{\rm pr}\, R}\big[A^{\Lambda \Sigma}_{ij}(x,ty,tz)\omega^{i}_{\Lambda}\wedge \omega^{j}_{\Sigma}\wedge \eta\big]=tA^{\Lambda \Sigma}_{ij}(x,ty,tz)(z^{i}_{\Lambda}\omega^{j}_{\Sigma}-z^{j}_{\Sigma}\omega^{i}_{\Lambda})\wedge \eta.
\end{gather*}
Using this in \eqref{Vhom2} we get
\[
h^{n,2}_{V}\omega =\left(\int_{0}^{1}tA^{\Lambda \Sigma}_{ij}[x,ty,tz]dt \right)\big(z^{i}_{\Lambda}\omega^{j}_{\Sigma}-z^{j}_{\Sigma}\omega^{i}_{\Lambda}\big)\wedge \eta.
\]
Apply this for $\omega =d_{V}K_{C}$.  We have (using the equality $d_V \omega^{i}_{\sigma}=0$),
\[
d_{V}K_{C}=(d_V \Pi_{i})\wedge \omega^{i}\wedge \eta +(d_V F^{\mu}_{i})\wedge\omega^{i}_{\mu}\wedge \eta=\Pi_{i,z^{j}_{\Sigma}}\omega^{j}_{\Sigma}\wedge \omega^{i}\wedge \eta+
F^{\mu}_{i,z^{j}_{\Sigma}}\omega^{j}_{\Sigma}\wedge\omega^{i}_{\mu}\wedge \eta.
\]
In this formula $\vert \Sigma \vert \geq 0$.  Combining this result with the previous calculation we get the complemental, ``pure non-Lagrangian'', term in the decomposition \eqref{Vdec}  of the form $K_{C}$:
\begin{gather}
h^{n,2}_{V}(d_V K_{C}))=\left( \int_{0}^{1}t\Pi_{i,z^{j}_{\Sigma}}(x,ty,tz)dt \right) \big(z^{j}_{\Sigma}\omega^i -y^i \omega^{j}_{\Sigma}\big)\wedge \eta\nonumber\\
\phantom{h^{n,2}_{V}(d_V K_{C}))=}{}
+ \left(\int_{0}^{1}tF^{\mu}_{i,z^{j}_{\Sigma}}(x,ty,tz)dt  \right)\big(z^{j}_{\Sigma}\omega^{i}_{\mu}-z^{i}_{\mu}\omega^{j}_{\Sigma}\big)\wedge \eta \nonumber\\
\phantom{h^{n,2}_{V}(d_V K_{C}))}{} = \left\{ \int_{0}^{1}t[(y^j \Pi_{i,y^j}[]+z^{j}_{\mu}\Pi_{i,z^{j}_{\mu}}[]+z^{j}_{\Lambda}\Pi_{i,z^{j}_{\Lambda}}[])-(
y^j \Pi_{j,y^i}[]-z^{j}_{\mu}F^{\mu}_{j,y^i}[])]dt\right\} \omega^{i}\wedge \eta \nonumber\\
\phantom{h^{n,2}_{V}(d_V K_{C}))=}{}+
\left\{ \int_{0}^{1}\! t[-y^j \Pi_{j,z^{i}_{\mu}}[]+y^j F^{\mu}_{i,y^j}[]+z^{j}_{\nu}F^{\mu}_{i,z^{j}_{\nu}}[]-z^{j}_{\nu}F^{\nu}_{j,z^{i}_{\mu}}[]+z^{j}_{\Lambda}F^{\mu}_{i,z^{j}_{\Lambda}}[]]dt\right\} \omega^{i}_{\mu}\wedge \eta \nonumber\\
\phantom{h^{n,2}_{V}(d_V K_{C}))=}{}+
\left\{ \int_{0}^{1}\!t[-y^j \Pi_{j,z^{i}_{\Lambda}}-z^{j}_{\nu}F^{\nu}_{j,z^{i}_{\Lambda}} ] dt \right\}\omega^{i}_{\Lambda} \wedge \eta . \label{nLag}
\end{gather}
In the last expression $\vert \Lambda \vert >1$ .

Thus, we get
\begin{proposition} Vertical splitting \eqref{Vdec} of a $(n,1)$-form $K_{C}=\Pi_{i}\omega^i \wedge \eta +F^{\mu}_{i}\omega^{i}_{\mu}\wedge \eta$ has the form
\begin{gather}
K_{C} =K_{C,{\rm Lag}}+K_{C,n{\rm Lag}}, \label{Kdec}
\end{gather}
 where the Lagrangian part $K_{C,{\rm Lag}}$ is given by \eqref{Lag}, while the non-Lagrangian part $K_{C,n{\rm Lag}}$ is given in \eqref{nLag}.
\end{proposition}

In the appendix we present several examples of the Lagrangian components (quasi-Lagrangian and the corresponding Euler--Lagrange equation) of several well known nonlinear equations having the form of balance equations (see Table~\ref{table1}).

\subsection{Properties of vertical decomposition}\label{section5.3}

Consider the bicomplex \eqref{Bik} over a domain $W\subset Y$ which is vertically star-shaped and covered by one f\/ibred chart, see \cite{Olver}.  An important special case of this situation is the case where $\pi:Y\rightarrow X$ is the vector bundle. Then, the vertical complex with the vertical dif\/ferential $d_V$ is $d_{V}$-exact (\cite[Theorem~5.58]{Olver}  or \cite[Chapter~4]{Anderson}): any $d_V$-closed $(n,r)$-form over $W$ is exact. In addition to this, vertical homotopy operator is def\/ined in the domain $W^{\infty}=\pi_{\infty 0}^{-1}(W).$

Denote by $Z^{n,r}$ the subspace of $d_{V}$-closed forms in  $\Omega^{n,r}$. Introduce the subspace $Q^{n,r}=h^{n,r+1}Z^{n,r+1}\subset \Omega^{n,r}$. In the decomposition $\omega =h^{n,r+1}d_{V}\omega +d_{V} h^{n,r}\omega$,
\[
h^{n,r+1}d_{V}\omega\in Q^{n,r},\qquad
d_{V} h^{n,r}\omega\in Z^{n,r}.
\]

\begin{proposition} \label{proposition5} Locally, in the domain $W^{\infty}$ over a vertically star-shaped and covered by one fibred chart domain $W\subset Y$ the following is true:
\begin{enumerate}\itemsep=0pt
\item[$1.$] The decomposition $(K$-decomposition$)$
\begin{gather}\label{(K)}
\Omega^{n,r}(W^{\infty})=Q^{n,r}(W^{\infty})\oplus  Z^{n,r}(W^{\infty}) \tag{$K$}
\end{gather}
is direct over the domain $W$.
\item[$2.$] The presentation of a $(n,r)$-form $\omega$
\begin{gather}
\omega =h^{n,r+1}d_{V}\omega +d_{V} h^{n,r}\omega \label{Pres}
\end{gather}
as the sum of terms in the decomposition \eqref{(K)} \textit{is unique}.
\item[$3.$] The mapping $V=h^{n,r+1}\circ d_V:\Omega^{n,r}\rightarrow \Omega^{n,r}$ is the projector: $V^2=V$.
\item[$4.$] The mappings
\[
Z^{n,r+1}\xrightarrow{h^{n,r+1}}  Q^{n,r}\qquad \text{and}\qquad
 Q^{n,r} \xrightarrow{d_V}Z^{n,r+1}
\]
are  linear isomorphisms inverse to one another.
\item[$5.$] For $\nu \in \Omega^{n,1}$, $h^{n,r}\nu \in Z^{n,r-1}$ iff $\nu \in Q^{n,r}$.
\item[$6.$] An $(n,1)$-form $\omega$ is anti-Lagrangian, $\omega =h^{n,r+1}d_{V}\omega \ \Leftrightarrow \ h^{n,r}\omega \in \pi^{*}\Omega^{n}(X)$.
\item[$7.$] Over the domain $W^{\infty}$ the following direct decomposition $(F$-decomposition$)$ at the level of functional is valid
\begin{gather}\label{(F)}
\mathcal{F}^{1}\big(W^{\infty}\big)=I\big(Q^{n,1}\big)\oplus \mathcal{E}\big(\Omega^{n,0}\big), \tag{$F$}
\end{gather}
where $\mathcal{E}(\Omega^{n,0})$ is the space of Lagrangian functional forms.
\end{enumerate}
\end{proposition}

\begin{proof} To prove the f\/irst statement, it is suf\/f\/icient to show that the intersection of the two subspaces in the decomposition \eqref{(K)} is zero.
Let $\omega \in Q^{n,r}\bigcap  Z^{n,r}(W^{\infty})$, i.e.
\begin{gather*}
\omega = h^{n,r+1}\sigma = d_{V}\lambda,
\end{gather*}
where $\sigma=d_{V}\sigma'$, $\sigma'\in \Omega^{n,r}$ and $\lambda \in \Omega^{n,r-1}$.  Applying $d_{V}$ to this decomposition of $\omega$ we get
$d_{V}\omega =0=d_{V}h^{n,r+1}\sigma$ and, therefore, $d_{V}h^{n,r+1} d_{V}\sigma'=0$. Using the decomposition~\eqref{Vdec} for $\sigma'$ we have $h^{n,r+1} d_{V}\sigma' =\sigma'-d_{V}h^{n,r}\sigma'$. Substituting this equality into the previous one, we get
\[
d_{V}h^{n,r+1} d_{V}\sigma'=d_{V}(\sigma'-d_{V}h^{n,r}\sigma')=d_{V}\sigma' =0.
\]
It follows from this that $\sigma'\in Z^{n,r}$ and that $\sigma=d_{V}\sigma'=0$. Therefore, $\omega =0$.

The second statement is another form of the f\/irst one.  The third statement follows from the second one if we apply~$V$ to the decomposition~\eqref{Pres}.

To prove 5th statement let $\nu \in Q^{n,r}$, then, $\nu =h^{n,r+1} d_{V}\nu +d_{V} h^{n,r}\nu$.  By unicity of such presentation, $d_{V}h^{n,r}\nu=0$.  This proves the 5th statement and part of statement~4.

Let $\sigma \in Z^{n,r+1}$, then $\sigma =d_{V}\nu, \nu \in  Q^{n,r}$. We have $h^{n,r+1}\sigma =h^{n,r+1}d_{V}\nu$ and
\[
d_V h^{n,r+1}\sigma =d_V h^{n,r+1}d_{V}\nu=d_V \big(1-d_V h^{n,r+1}\big)\nu=d_V \nu =\sigma.
\]
Combining this with the proven statement $\nu =h^{n,r+1}d_{V}\nu$ we f\/inish the proof of forth statement.

Next statement follows from the fact that ${\rm Ker}\, (d_{V})$ in $\Omega^{n,0}$ coincides with $\pi^{*}\Omega^{n}(X)$.

To prove the last statement that the sum \eqref{(F)} is direct we use the fact that the homotopy operator $h_V$ and the horizontal dif\/ferential $d_H$ anticommute: If
 $\omega \in Q$ and if  $I(\omega)=I(d_{V} L\eta)$ for some function $L$, then $\omega -d_{V}L\eta =d_{H}\beta$ for some $\beta \in \Omega^{n-1,1}$. Applying decomposition \eqref{(K)} to the form $d_H \beta$: $d_{H}\beta =h^{n,2}d_V d_H \beta +d_{V}h^{n,1} d_{H}\beta$, substituting it into the expression  $\omega =d_{V}L\eta +d_{H}\beta$ for $\omega$ and using unicity in the decomposition \eqref{(K)} we see that $\omega =h^{n,2} d_{V}d_{H}\beta =d_{H}h^{n-1,2}d_{V}\beta $.  As a result, $I(\omega)=0$.
\end{proof}

\begin{remark} I am grateful to the the referee for the remark that if the conf\/igurational bundle $\pi:Y\rightarrow X$ is vector bundle, decomposition~\eqref{Vdec} and~\eqref{Pres} are globally def\/ined and depend only on the structure of the vector bundle $\pi$.
\end{remark}

\begin{remark} Projectors $I$ and $V$ do not commute. It follows from the fact that the forms in the image of $I$  are of the type $P_{i}\omega^{i}\wedge \eta$ while the forms in the image of $V$ have also terms $A^{\Lambda}_{i}\omega^{i}_{\Lambda}\wedge \eta$ with $\vert \Lambda >0$.
\end{remark}

\begin{remark}
It would be interesting to know, if the restriction of $h^{n,r}_{V}$ to $Q^{n,r}$ is the monomorphism?  Is it onto the $Z^{n,r-1}$?
\end{remark}

\subsection[$(n,1)$-forms $K$ with zero Lagrangian part]{$\boldsymbol{(n,1)}$-forms $\boldsymbol{K}$ with zero Lagrangian part}\label{section5.4}

Here we study the $(n,1)$-forms $K_C$ which have zero Lagrangian part.

We will be using the following simple lemma.

\begin{lemma}\label{lemma1} Let $f(u^i )$ be a function of $m$ variables $u^i$, analytical in a connected domain $D\subset {\mathbb R}^m$ containing the origin $0$. Let $s\in \mathbb{Z}$ be an integer number. If
\[
\int_{0}^{1}t^s f(tu)dt=c-{\rm const}\ \mbox{$($respectively   zero$)$}
\]
in an open subset $O\subset D$ containing the origin, then $f(u)=(s+1)c-{\rm const}$ $($respectively zero$)$ for all $u\in D$.
\end{lemma}

\begin{proof}
Present function $f$ by its Taylor series centered at the origin $0$;
\begin{gather}
f(u)=\sum_{I, \vert I\vert \geq 0}c_{I}u^{I},\label{Ser}
\end{gather}
where $u^I=u^{1\ i_{1}}\cdots u^{m\ i_m}$ is the monom corresponding to the ordered multiindex $I$. This Taylor series absolutely converges in some disc $\Vert u\Vert <R$, where $R>0$. Rearrange this series collecting terms $c_{I}u^{I}$ with $\vert I\vert =k$:
\[
f(u)=\sum_{k=0}^{\infty} u^{\{k\}},
\]
where $u^{\{k\}}$ is the homogeneous part of series \eqref{Ser} of degree $k$. This series also absolutely converges in the disc $\Vert u\Vert <R$. We have, now
\[
f(tu)=\sum_{k=0}^{\infty}u^{\{k\}}t^k.
\]
As a result,
\begin{gather}\label{Dseries}
\int_{0}^{1}t^s f(tu)dt=\sum_{k=0}^{\infty}u^{\{k\}}\int_{0}^{1}t^{s+k}dt=\sum_{k=0}^{\infty}\frac{1}{k+s+1}u^{\{k\}}=\sum_{I, \vert I\vert \geq 0}\frac{1}{\vert I\vert +s+1}c_{I}u^{I}.
\end{gather}
Last series has the same radius of convergence as the series \eqref{Ser}.  Under the condition of lemma, series \eqref{Dseries} equal constant~$c$ (respectively  zero) for all $u$ in the disc $\Vert u\Vert <R$. This series presents an analytical function and is equal to the constant (respectively zero) in the disc $\Vert u\Vert <R$.  Therefore, series \eqref{Dseries} have all but the zeroth (respectively all) Taylor coef\/f\/icients equal zero.  Then, the same is true for the series~\eqref{Ser} for $f(u)$. Being equal constant (being identically zero) in an open subset of its domain, function~$f$ is the constant (identically zero) in the whole connected component of its domain containing zero.  This proves the lemma.
\end{proof}

Applying this lemma to the quasi-Lagrangian $\tilde L$ given by \eqref{qL} presented as the sum of homogeneous components
\begin{gather*}
\tilde L  =\int^{1}_{0}\sum_{k=0}^{\infty}\big[y^i \Pi_{ik}(x,y,z)  +z^{i}_{\mu}F^{\mu}_{ik}(x,y,z)\big]t^k dt
\end{gather*}
we prove the f\/irst statement in the following

\begin{proposition} Let $K_{C}=\Pi_{i}(x,y^i, z^{i}_{\mu})\omega^i\wedge \eta+F^{\mu}_{i}(x,y^i, z^{i}_{\mu})\omega^{i}_{\mu}\wedge \eta$ be a $(n,1)$-form analytical by vertical variables $y^i$, $z^{i}_{\mu}$ in a connected domain $D\subset {\mathbb R}^{m+mn}$ containing the origin.  Then,
\begin{enumerate}\itemsep=0pt
\item[$1.$] The quasi-Lagrangian $\tilde L$ corresponding to the form $K_{C}$ is a function of $x$ only $($identically equals zero$)$ if and only if
\begin{gather}
y^i \Pi_{i}(x,y,z)   +z^{i}_{\mu}F^{\mu}_{i}(x,y,z)=\phi(x)\ \ \mbox{$($respectively, $=0)$}. \label{cond}
\end{gather}
\item[$2.$] If this equality holds, then $\tilde L=0$.
\end{enumerate}
\end{proposition}

\begin{proof} Second statement follows from the f\/irst and the fact that the analytical
function in the left side of~\eqref{cond} vanish when $y^i=z^{i}_{\mu}=0$.
\end{proof}

\section{Balance systems of zero order}\label{section6}

\subsection[Balance systems whose quasi-Lagrangian $\tilde L$ has zero Euler-Lagrange form]{Balance systems whose quasi-Lagrangian $\boldsymbol{\tilde L}$ has zero Euler--Lagrange form}\label{section6.1}

Here we look at the case where  Lagrangian part does not enter the functional form of $K_{C}$ (pure non-Lagrangian case).

We consider here only balance systems of order zero (called the RET systems in continuum thermodynamics~\cite{MullerRuggeri}).  Thus, the density-f\/luxes $F^{\mu}_{i}$ and the source terms $\Pi_{i}$ depend on the f\/ields $y^i$ and, possibly, on $x^\mu$.

Lagrangian functions $L$ whose $(n,1)$-form $d_{V}L \wedge \eta$ is of order zero are linear by the derivatives: $L=L_{0}(x,y)+z^{i}_{\mu}L^{\mu}_{i}(x,y)$ and the corresponding Euler--Lagrange systems are quasilinear systems of the f\/irst order
\begin{gather}
\big(L^{\mu}_{i,y^k}-L^{\mu}_{k,y^i}\big)y^{k}_{,x^\mu}+\big(L_{0,y^i}-L^{\mu}_{i,x^\mu}\big)=0,\qquad i=1,\ldots,m. \label{LagC}
\end{gather}
Consider now an arbitrary form $K=\Pi_{i}\omega^{i}\wedge \eta +F^{\mu}_{i}\omega^{i}_{\mu}\wedge \eta$ of order zero. Quasi-Lagrangian $\tilde L$ corresponding to this form is
\[
\tilde L =\int_{0}^{1}\big[y^k\Pi_{k}(x,ty)+z^{k}_{\nu}F^{\nu}_{k}(x,ty) \big]dt.
\]
This function depends linearly on the jet variables.

Applying the Euler operator $I$ to the form $d_{V}({\tilde L}\eta)$ we get
\[
I d_{V}({\tilde L}\eta)=\left({\tilde L}_{,y^i}-d_{\mu}\frac{\partial \tilde L}{\partial z^{i}_{\mu}} \right)\omega^i\wedge \eta.
\]
In the zero order case,
\begin{gather*}
{\tilde L}_{,y^i}-d_{\mu}\frac{\partial \tilde L}{\partial z^{i}_{\mu}} =\int_{0}^{1}\left\{\Pi_{i}[]+ty^k \Pi_{k,y^i}[]+tz^{k}_{\nu}F^{\nu}_{k,y^i}[] \right\}dt-d_{\mu}\int_{0}^{1}F^{\mu}_{i}[]dt \\
\phantom{{\tilde L}_{,y^i}-d_{\mu}\frac{\partial \tilde L}{\partial z^{i}_{\mu}}}{} =
\int_{0}^{1}\left\{\Pi_{i}[]+ty^k \Pi_{k,y^i}[]+tz^{k}_{\nu}F^{\nu}_{k,y^i}[] \right\}dt-\int_{0}^{1}\left\{ F^{\mu}_{i,\mu}[]+tz^{k}_{\mu}F^{\mu}_{i,y^k}[]\right\}dt.
\end{gather*}
In the homogeneous case, where there is no no explicit dependence of $F^{\mu}_{i}$ on $x^\mu$, the f\/irst term in the second integral vanishes and we get
\begin{gather*}
{\tilde L}_{,y^i}-d_{\mu}\frac{\partial \tilde L}{\partial z^{i}_{\mu}} =\int_{0}^{1}\left\{\Pi_{i}[]+ty^k \Pi_{k,y^i}[] \right\}dt+\int_{0}^{1}\left\{tz^{k}_{\nu}F^{\nu}_{k,y^i}[]- tz^{k}_{\mu}F^{\mu}_{i,y^k}[]\right\}dt \\
\phantom{{\tilde L}_{,y^i}-d_{\mu}\frac{\partial \tilde L}{\partial z^{i}_{\mu}}}{} =
\int_{0}^{1}\left\{\Pi_{i}[]+ty^k \Pi_{k,y^i}[] \right\}dt+z^{k}_{\mu}\left\{\left(\int_{0}^{1}F^{\mu}_{k}[]dt \right)_{,y^i}- \left(\int_{0}^{1} F^{\mu}_{i}[]dt\right)_{,y^k}\right\}.
\end{gather*}
The variables $z^{k}_{\mu}$ are independent and are present only as factors in the second term of last line.  so, ${\tilde L}_{,y^i}-d_{\mu}\frac{\partial \tilde L}{\partial z^{i}_{\mu}} =0$ if and only if both terms -- one containing $\Pi_{i}$ and the others containing $F^{\mu}_{i}$ all vanish.  In particular, we have to have
\[
\left(\int_{0}^{1}F^{\mu}_{k}[]dt \right)_{,y^i}= \left(\int_{0}^{1} F^{\mu}_{i}[]dt\right)_{,y^k},\qquad \forall \; i,k.
\]
This condition has the form of mixed derivative test and as a result, Euler--Lagrange equations for Lagrangian $\tilde L$ vanishes if and only if the following conditions holds:
\begin{gather*}
\int_{0}^{1}\left\{\Pi_{i}[]+ty^k \Pi_{k,y^i}[] \right\}dt=\frac{\partial}{\partial y^i}\int_{0}^{1}y^k\Pi_{k}[]dt =0,
\end{gather*}
locally, for some functions~$G^{\mu}(y^i)$,
\begin{gather*}
\int_{0}^{1}F^{\mu}_{k}(ty)dt =\frac{\partial G^{\mu}}{\partial y^k},\qquad \mu=1,\ldots,n,\quad k=1,\ldots,m.
\end{gather*}
Using Lemma~\ref{lemma1} we show that in the analytical case the f\/irst condition is equivalent to the condition $y^k\Pi_{k}(x,y)=c(x)$ and, due to the analyticity by vertical variables, to the equality $y^k\Pi_{k}(x,y)=0$. More than this, in the analytical case we can rearrange terms of Taylor series for the functions~$F^{\mu}_{k}$ and $G^{\mu}(y)$ collecting together monoms $y^I=y^{1 i_{1}}\cdots y^{m\ i_{m}}$ with the same value of $\vert I\vert=k$:
\begin{gather}
F^{\mu}_{k}=\sum_{p}F^{\mu}_{kp},\qquad G^{\mu}=\sum_{l=0}^{\infty}G^{\mu}_{l}. \label{Hom}
\end{gather}
Substituting this into the second condition and using Lemma~\ref{lemma1}, we write it in the form
\[
\sum_{l=0}^{\infty}\frac{\partial G^{\mu}_{l}}{\partial y^k}=\sum_{p}F^{\mu}_{kp}(y)\int_{0}^{1}t^{p}dt =\sum_{p}\frac{1}{p+1}F^{\mu}_{kp}(y).
\]
Equating homogeneous terms of the same order in both parts we get
\[
\frac{1}{p+1}F^{\mu}_{kp}(y)=\frac{\partial G^{\mu}_{p+1}}{\partial y^k},\qquad p=0,1,\ldots .
\]
Multiplying by $(p+1)$ and taking summation by $p$ we get the formal equality
\[
F^{\mu}_{k}=\sum_{p=0}^{\infty}F^{\mu}_{kp}(y)=\sum_{p=0}^{\infty} \big((p+1)G^{\mu}_{p+1}\big)_{, y^k}=\frac{\partial }{\partial y^k}\left(\sum_{p=0}^{\infty} (p+1)G^{\mu}_{p+1}\right).
\]
It is easy to see that the series on the right side has the same radius of convergence as the corresponding series in \eqref{Hom}.  Denote obtained analytical function by
\[
{\tilde G}^{\mu}(y)=\sum_{p=0}^{\infty} (p+1)G^{\mu}_{p+1}(y).
\]
As a result, we have proved the following
\begin{proposition} Let the balance system \eqref{Bsys} be analytical and of order zero $($RET case$)$.  Then the Lagrangian part of this balance system in the functional form splitting \eqref{(F)} vanishes, or, in other terms, $I(K_{C})\in I(Q^{n,1})$ if and only if the following conditions are fulfilled:
\begin{enumerate}\itemsep=0pt
\item[$1)$] $y^k\Pi_{k}(x,y)=c-\text{\rm const}$,
\item[$2)$]  Godunov condition $($see below  Definition~{\rm \ref{definition1})}: the form $F^{\mu}_{i} \omega^{i}\wedge \eta_{\mu}$ is vertically closed, i.e.\ locally, $F^{\mu}_{i} \omega^{i}\wedge \eta_{\mu}=d_{V}(G^{\mu}\eta_{\mu})$ for some analytical functions $\tilde{G}^{\mu}(y^i)$, $\mu=1,\ldots, n$.  As a~result,
\begin{gather*}
F^{\mu}_{k}(y) =\frac{\partial \tilde{G}^{\mu}}{\partial y^k},\qquad \mu=1,\ldots,n,\quad  k=1,\ldots,m.
\end{gather*}
\end{enumerate} \label{proposition7}
\end{proposition}

\begin{remark} Notice that the Godunov condition has the form $d_{V}F^{\mu}_{i}\omega^{i}\wedge
\eta_{\mu}=0$, is covariant and, therefore, globally well def\/ined. The second condition: $y^i\Pi_{i}=0$ can be written, using the Liouville vector f\/ield $R=y^i\partial_{y^i}$, in the form $i_{R}K_{C}=0$. Thus, this condition is globally well def\/ined on the vector bundles $\pi:Y\rightarrow X$ where the transition transformations are linear on the f\/ibers of $\pi$.
\end{remark}

\subsection[Systems of Godunov and the decomposition of the $(n,1)$-forms of order zero]{Systems of Godunov and the decomposition of the $\boldsymbol{(n,1)}$-forms\\ of order zero}\label{section6.2}

In 1961, S.~Godunov \cite{Godunov1961} def\/ined an interesting class of balance systems of the form
\begin{gather}
\sum_{\mu=0}^{\mu=3}\frac{d}{d x^\mu}\left(\frac{\partial G^\mu}{\partial y^i} \right)=0,\qquad i=1,\ldots, m,\label{God}
\end{gather}
where $G^{\mu}(y)$ are some functions of the dynamical f\/ields~$y^i$. These systems are generalizations of symmetrical hyperbolic systems of Friedrichs. Later on, S.~Godunov and his collaborators used these systems as a~starting point for constructing Galilean invariant and thermodynamically compatible models of a~large class of physical systems~\cite{GodunovGordienko,Godunov2003}. This was done by adding terms depending on the derivatives~$z^{i}_{\mu}$, to the f\/luxes $F^{A}_{i}$, $A=1,2,3$ and introducing the source terms~$\Pi_{i}$. At the same time, G.~Boillat and T.~Ruggeri proved that the Godunov systems~\eqref{God} present the canonical form of a RET (zero order) balance system with the entropy balance if the entropy density is convex (condition guaranteeing the thermodynamical stability) \cite[Chapter~3, Section~2]{MullerRuggeri}.

Thus, it seems natural to introduce
\begin{definition} \label{definition1} A balance system \eqref{Bsys} of order zero and the corresponding $(n,1)$-form $K$ will be called Godunov type system if the density-f\/lux part of the system has the form
\[
F^{\mu}_{i}=\frac{\partial G^\mu}{\partial y^i}
\]
for a functions $G^{\mu}\in C^{\infty}(J^{k}(\pi)).$
\end{definition}

 Thus, the result of Proposition~\ref{proposition7} can be formulated as follows:
\begin{corollary} An analytical balance system of zero order is pure non-Lagrangian if and only if it is of Godunov type and its source part $\Pi_{i}\omega^{i}\wedge \eta$ satisfies to the condition
\[
y^k\Pi_{k}(x,y)=0.
\]
\end{corollary}

Next we project this result to the complex of functional forms (most right column in~\eqref{Bik}), where $\delta_{V}=I\circ d_V$ is the induced vertical dif\/ferential. Using the vertical splitting \eqref{(F)} of functional forms (see also~\cite[Chapter~4A]{Anderson}) we get the following

\begin{corollary} Let the domain $W\subset Y$ satisfies to the conditions of Proposition~{\rm \ref{proposition5}}.  Then, any analytical functional form $K\in \mathcal{F}^1$ of order zero in the domain $W$ is the sum
\begin{gather*}
I(K)=G_{K}+\mathcal{E}_{\tilde{L}_{K}}
\end{gather*}
 of the functional forms generated by:
 form of Godunov type
\[ G_{K}=\big[(\Pi_{i}-{\tilde L}_{K,y^i})\omega^i +(F^{\mu}_{i}-{\tilde L}_{K,z^{i}_{\mu}})\omega^{i}_{\mu}\big]\wedge \eta \]
and the Euler form
\[\mathcal{E}_{\tilde{L}}=d_{V}{\tilde L}_{K} \wedge \eta\]
 for the quasi-Lagrangian ${\tilde L}_{K}$. This decomposition is unique $($in the domain $W)$.
\end{corollary}

\begin{proof}
Present the $(n,1)$-form $K$ as the sum $K=K_{1}+d_{V}({\tilde{L}_K} \eta)$ as in~\eqref{Kdec}.   Component $K_{1}\in Q^{n,1}$ in this representation has the property that ${\tilde{L}_{K_{1}}}\in C^{\infty}(X)$ and, therefore, $\mathcal{E}_{{\mathcal{L}_{K_{1}}}}=0$.  From the previous corollary it follows that $K_{1}$ is of \emph{Godunov type}. Then, since $I^2=I$,  $K=I(K)=I(K_1 )+\mathcal{E}_{{\tilde{L}_K}}.$ Now we rename $I(K_1 )=G_{K}$ and the proof is complete.
\end{proof}

 Comparing systems of equations corresponding to the Lagrangian component $d_{V}({\tilde L}\eta)$ and the Godunov component of a $(n,1)$-form $K=\Pi_{i}\omega^i \wedge \eta +F^{\mu}_{i}\omega^{i}_{\mu}\wedge \eta$ with $\Pi_{i}$, $F^{\mu}_{i}\in C^{\infty}(Y)$, we see that the tensor $A^{\mu}_{ij}$ of the principal parts $A^{\mu}_{ij}(x,y)\partial_{y^{j}_{,x^\mu}}$ (of $i$-th equation) is symmetrical (by~$ij$) for the Godunov component of the balance system and anti-symmetrical (see~\eqref{LagC}) for the Lagrangian component.

Let now $K=\Pi_{i}(x,y)\omega^{i}\wedge \eta +F^{\mu}_{i}(x,y)\omega^{i}_{\mu}\wedge \eta$ be a $(n,1)$-form of order zero of the type considered here. Let $\tilde L={\tilde L}_{0}(x,y)+z^{i}_{\mu}{\tilde L}^{\mu}_{i}(x,y)$ be the corresponding quasi-Lagrangian.  Then
${\tilde L}^{0}_{i}=\int_{0}^{1}F^{0}_{i}(x,ty)dt.$ Godunov component $G_{K}$ of the form $K$ has the functions ${\tilde F}^{\mu}_{i}=F^{\mu}_{i}-{\tilde L}_{K,z^{i}_{\mu}}=F^{\mu}_{i}-\int_{0}^{1}F^{\mu}_{i}(x,ty)dt$ as its densities ($\mu=0$) and f\/luxes ($\mu=1,2,3$).  Matrices
\[
{\tilde F}^{\mu}_{i,y^j}=F^{\mu}_{i,y^j}-\int_{0}^{1}tF^{\mu}_{i,y^j}(x,ty)dt
\]
are symmetric by $(i,j)$ (see Def\/inition~\ref{definition1}). As a result, the system of balance equations corresponding to the form $G_{K}$ is the f\/irst order system of PDEs $M^{\mu}_{ij}(x,y)y^{j}_{,\mu}=b_{i}(x,y)$ \emph{with symmetrical matrices} $M^{\mu}_{ij}$.  If the matrix $M^{0}_{ij}={\tilde F}^{0}_{i,y^j}$
\emph{is positive definite}, this system is the symmetrical hyperbolic by Friedrichs~\cite{MullerRuggeri}.  Thus, we get
\begin{corollary} Let $K=\Pi_{i}(x,y)\omega^{i}\wedge \eta +F^{\mu}_{i}(x,y)\omega^{i}_{\mu}\wedge \eta$ be an analytical by $y^i$ $(n,1)$-form of order zero and let the
symmetrical matrix $M^{0}_{ij}={\tilde F}^{0}_{i,y^j}$ \emph{is positive definite}.  Then the system of balance equations corresponding to the Godunov component $G_{K}$ of the form $K$ is the symmetrical hyperbolic system by Friedrichs.
\end{corollary}

We f\/inish this section with two examples of calculation of quasi-Lagrangian for two balance systems -- the system of elasticity equations for a hyperalastic body and the reduced system of equations  of 2-dimensional ideal plasticity.

\subsection{Example: hyperelasticity}\label{section6.3}

Consider the system of equations of hyperelasticity  \cite[Section~1.6]{Serre}.

Let $M^{3}$ be a material manifold -- a connected open 3-dimensional manifold endowed with the (reference) Riemannian metric $g_{0}$. Local coordinates in $M$ will be denoted by $X^I$, $I=1,2,3$, time~$t$ by~$X^0$.  A deformation history of the body $M$ is a~time dependent dif\/feomorphic embedding $\phi:M\rightarrow {\mathbb E}^3$ of $M$ into the Euclidian 3-dimensional (physical) space~${\mathbb E}^3$ endowed with the metric $h$.  Local coordinates in ${\mathbb E}^3$ will be denoted by $x^i$, $i=1,2,3$.  As the dynamical variables of Elasticity we choose the velocity vector f\/ield over $\phi$: $\mathbf{v}=\frac{\partial \phi}{\partial X^0}$ and the deformation gradient~$\mathbf{F}$, $F^{i}_{I}=\frac{\partial \phi^i}{\partial X^I}$ -- $(1,1)$-tensor over the mapping~$\phi$.

Constitutive properties of the body are determined by the strain energy function $w=w(F^{i}_{\mu})$. Strain energy $w$ depends on the material metric $g_{0}$ and the physical Euclidian metric~$h$~\cite{MarsdenHughes}.

Elasticity equations have the form of the balance system of order zero, where the f\/irst equation is the linear momentum balance law and the second one -- compatibility condition,
\begin{gather}
\partial_{X^0}v_i +\partial_{X^I}\left(\frac{\partial w}{\partial F^{i}_{I}}\right)=b_{i}(X),\qquad
\partial_{X^0}F^{i}_{I}-\partial_{X^I}\big(\delta^{i}_{k}v^k\big)=0.
  \label{El}
\end{gather}

Dynamical f\/ields are $y^i=v^i$, $i=1,2,3$ and $y^{aA}=F^{a}_{A}$, $a,A=1,2,3$. Jet variables will be denoted by $z^{i}_{\mu}$, $i=1,2,3$, $\mu=0,1,2,3$, corresponding to the derivatives of velocity components and $z^{aA}_{\mu}$, $a,A=1,2,3$, $\mu=0,1,2,3$, corresponding to the derivatives of components of the deformation gradient.  Basic contact forms in the second jet bundle $J^{2}(\pi)$ are
\begin{gather*}
\omega^i=dv^i -z^{i}_{\nu}dX^\nu,\qquad
\omega^{aA}=dF^{a}_{A}-z^{aA}_{\nu}dX^\nu,\\
\omega^{i}_{\mu}=dv^{i}_{,\mu}-z^{i}_{\mu \nu}dX^\nu,\qquad
\omega^{aA}_{\mu}=dF^{a}_{A,\mu}-z^{a}_{A,\mu \nu}dX^\nu.
\end{gather*}
The form $K_{C}$ in this case is
\[
K_{C}=(b_{i}\omega^i \wedge \eta)+\big(v_{i}\omega^{i}_{0}+w_{,F^{a}_{A}}\omega^{aA}\big)\wedge \eta +\big(F^{a}_{A}\omega^{aA}_{0}-\delta^{B}_{A}v^A \omega^{aA}_{B}\big)\wedge \eta.
\]
To calculate the quasi-Lagrangian ${\tilde L}=\int_{0}^{1}[y^i\Pi_{i}[]+z^{i}_{0}F^{0}_{i}[]+z^{i}_{A}F^{A}_{i}[]]dt$ we write
\[
\int_{0}^{1}[y^i\Pi_{i}]dt=v^i b_{i},
\]
which has the meaning of mechanical power of the bulk force $b_{i}dx^i$ (we assume that $b_j (X)$ are functions of $X$ only).

Next, using the compatibility equation, we get for the terms corresponding to the elasticity equations
\begin{gather*}
\int_{0}^{1}\big(z^{i}_{0}F^{0}_{i}[]+z^{i}_{A}F^{A}_{i}[]\big)dt=\int_{0}^{1}\big(z^{i}_{0}tv_i+z^{i}_{A}w_{,F^{i}_{A}}[]\big)dt\sim    \frac{1}{2} v_{i}\partial_t v^i +v^{i}_{,x^A}\int_{0}^{1}w_{,F^{i}_{A}}[]dt \\
\qquad{} = \partial_{X^0}\left(\frac{\Vert \mathbf{v}\Vert^2}{2} \right)+(\partial_{X^0}F^{i}_{A})\partial_{F^{i}_{A}}\int_{0}^{1}t^{-1}w[]dt=\partial_{X^0}\left( \frac{\Vert \mathbf{v}\Vert^2}{2} +\int_{0}^{1}t^{-1}w[]dt\right).
\end{gather*}
We recognize the time derivative of the energy density of elastic body (sum of kinetic energy and strain energy). Finally, using $F^{a}_{B,0}=v^{a}_{,X^B}$, we get
\begin{gather*}
\int_{0}^{1}\big[ z^{aB}_{0}F^{0}_{aB}+z^{aB}_{A}F^{A}_{aB}\big] dt=\int_{0}^{1}\big[z^{aB}_{0}F^{a}_{B}t+z^{aB}_{A}\delta^{A}_{B}v^a t\big] dt  \sim \frac{1}{2}\big(  F^{a}_{B,0}F^{a}_{B}+F^{a}_{B,X^B}v^a\big)  \\
\qquad{} =
\frac{1}{2}\big(v^{a}_{,B}F^{a}_{B}+v^a F^{a}_{B,x^B}\big)=\frac{1}{2}\partial_{X^B}\big( v^a F^{a}_{B} \big).
\end{gather*}
Combining we get (partial derivatives here coincide with the total derivatives)
\begin{gather*}
{\tilde L}=v^i b_{i}+d_{X^0}\left( \frac{\Vert \mathbf{v}\Vert^2}{2} +\int_{0}^{1}t^{-1}w[]dt\right)+d_{X^B}\left(\frac{1}{2} v^a F^{a}_{B} \right).
\end{gather*}
Thus, the quasi-Lagrangian is the sum of the trivial Lagrangian and the source term that generates the right side in the elasticity equation~\eqref{El}. Notice that the trivial part of $\tilde L$ has the form of the left side of a balance law similar to the energy balance for the elastic body.

\subsection{Example: 2-dimensional ideal plasticity}\label{section6.4}

System of equations described the plane strained state of the medium with von Mises conditions can be transformed, by change of variables (see~\cite{SenashovVinogradov}), to the system of balance equations
\begin{gather}
u_{,\xi}+\frac{1}{2}v=0,\qquad
v_{,\eta}+\frac{1}{2}u=0
  \label{Plas}
\end{gather}
for the functions $y^1=u$, $y^2=v$ of variables $x^0=\xi$, $x^1=\eta$. For this balance system, $F^{0}_{1}=u$, $F^{1}_{1}=0$, $F^{0}_{2}=0$, $F^{1}_{2}=v$, $\Pi_{1}=-\frac{1}{2}v$, $\Pi_{2}=-\frac{1}{2}u$.

The quasi-Lagrangian of this system is
\begin{gather*}
\tilde L =\frac{1}{2}\left[\partial_{\xi}\frac{u^2}{2}+\partial_{\eta}\frac{v^2}{2} \right]-uv.
\end{gather*}
Thus, in this case the quasi-Lagrangian is also sum of total divergence and of the expression producing the source terms in the balance system~\eqref{Plas}.

\section{Conclusion}\label{section7}

This works shows that the general systems of $m$ balance equations \eqref{Bsys} for $m$ dynamical f\/ields $y^i$ represent the natural type of systems of PDEs appearing in the framework of variational bicomplex. Any such system is associated with the $(n,1)$-form $K_{C}=\Pi_{i}\omega^{i}\wedge \eta +F^{\mu}_{i}\omega^{i}_{\mu}\wedge \eta \in \Omega^{n,1}(J^{\infty}(\pi))$.  It is proven that the vertical homotopy formula delivers, in appropriate domains or, for vector bundles, globally, the unique representation of $K_{C}$ as the sum of the Lagrangian term and the ``pure non-Lagrangian'' form.  The corresponding splitting of the functional forms follows from the decomposition of $(n,1)$-forms.  It is shown that for zero order forms $K_{C}$, whose quasi-Lagrangian $\tilde L$ is trivial ($I(d_{V}{\tilde L}\eta)=0$), the functional form $I(K_{C})$ is of the type introduced by S.~Godunov in 1961 \cite{Godunov1961}, and, later on, identif\/ied as the canonical (symmetrical hyperbolic) form of balance system in the Rational Irreducible Thermodynamics, \cite{MullerRuggeri}.

    In the continuation of this work we will study the decomposition~\eqref{Kdec} of the functional forms into the Lagrangian and non-Lagrangian parts for the balance systems of order one.  Another natural direction of  future work is to study the question (suggested by the referee):  is it possible to obtain a global version of the splitting \eqref{(F)}?  Finally, it would be interesting to extend results obtained for the case of analytical constitutive relations (density/f\/lux and source terms $F^{\mu}_{i}$, $\Pi_{i}$) to the  smooth ($C^{\infty}$) constitutive relations.

\vspace{-2mm}

\appendix

\section{Examples of vertical decomposition}\label{appendix}

In this appendix we present a few examples of the Lagrangian component of several well known nonlinear equations having the form of balance equations.  As these examples show, the terms in second order spacial derivatives arrive from the Lagrangian component.  The source terms in examples of Subsections~\ref{section6.3},~\ref{section6.4} also come from the Lagrangian component.  The same is true for the source term in the last equation below. It is possible that observations of this kind will be useful for the construction of thermodynamically compatible balance systems for dif\/ferent physical systems (comp.~\cite{GodunovGordienko,Godunov2003}).
\vspace{-3mm}

\begin{table}[h!]\centering\footnotesize
\caption{}\label{table1}
\vspace{1mm}
\begin{tabular}{|l|c|c|c|}
  \hline
  Name  & Equation  & quasi-Lagrangian $\tilde L$  & Lagr. part of eq. \tsep{1pt}\bsep{1pt}\\ \hline
\tsep{5pt}\bsep{2pt} $\!\!\begin{array}{@{}l@{}}\mbox{conser-} \\ \mbox{vation law}\end{array}$ & $\partial_{t}u+\partial_{x}C(u)=0$ &  $d_{t}\left(\frac{u^2}{4}\right)+d_{x}\left(\int_{0}^{1}\int_{0}^{u}C(ty)dy dt \right)$ & $0=0$
  \\ \hline
\tsep{7pt}\bsep{4pt} $\!\!\begin{array}{@{}l@{}}\mbox{KdV} \\
\mbox{equation}\end{array}$ & $\partial_{t}u+\partial_{x}\big(3u^2 +u_{xx}\big)=0$ & $d_{t}\left(\frac{u^2}{4}\right)+d_{x}\left(\frac{u^{2}_{,x}}{4}+\frac{u^3}{3}\right)$ & $0=0$ \\ \hline
\tsep{5pt}\bsep{2pt} $\!\!\begin{array}{@{}l@{}}\mbox{Burgers}\\
\mbox{equation}\end{array}$ & $\partial_{t}u-\partial_{x}\left(\frac{u^2}{2}+u_{x}\right)=0$ & $d_{t}\frac{u^2}{4}+d_{x}\left(-\frac{u^3}{18}\right)-\frac{u^{2}_{,x}}{2}$ & $u_{xx}=0$ \\ \hline
\tsep{5pt}\bsep{2pt} $\!\!\begin{array}{@{}l@{}}\mbox{Filtration}\\
\mbox{equation \cite{Krasil'shchikVinogradov}}\end{array}$ & $(u-\Delta u)_{,t}-(u_{,x^A})_{,x^A}=0$ &
$\left(\frac{u^2}{4}+\frac{\Vert \nabla u\Vert^2}{4} \right)_{,t}-\left(\frac{u_{,t}u_{,x^A}}{2}\right)_{,x^A}+\frac{\Vert \nabla u\Vert^2}{2}$ & $\Delta u=0$ \\ \hline
\tsep{5pt}\bsep{2pt} $\!\!\begin{array}{@{}l@{}}\mbox{cubic Schr\"{o}-}\\
\mbox{dinger equation}\end{array}$ & $iu_{,t}+\Delta u +\alpha \vert u\vert^2 u=0$ & $d_{t}(iu^2)+\frac{1}{2}\Vert \nabla u\Vert^2-\frac{\alpha}{4}u\vert u\vert^2$ & $\Delta u +\frac{\alpha}{4} \vert u\vert^2 u=0$ \\ \hline
\end{tabular}
\vspace{-3mm}
\end{table}

\subsection*{Acknowledgments} I would like to thank the second referee for his recommendations and comments. They allowed me to correct and/or clarify the formulations of some results and their proofs.

\pdfbookmark[1]{References}{ref}
\LastPageEnding

\end{document}